\documentstyle[preprint,aps]{revtex}

\begin{document}
\title{Fluctuations in the presence of fields\\
-Phenomenological Gaussian approximation and a new class of thermodynamic
inequalities- }
\author{S. Dumitru*, A. Boer**}
\address{Department of Physics, ''Transilvania'' University, B-dul Eroilor 29, R-2200%
\\
Brasov, Romania\\
e-mails: *) s.dumitru@unitbv.ro; **) boera@unitbv.ro }
\date{\today}
\maketitle

\begin{abstract}
The work approaches the study of the fluctuations for the thermodynamic
systems in the presence of the fields. The approach is of phenomenological
nature and developed in a Gaussian approximation. The study is exemplified
on the cases of a magnetizable continuum in a magnetoquasistatic field, as
well as for the so called discrete systems. In the last case one finds that
the fluctuations estimators depends both on the intrinsic properties of the
system and on the characteristics of the environment. Following some earlier
ideas of one of the authors we present a new class of thermodynamic
inequalities for the systems investigated in this paper. In the case of two
variables the mentioned inequalities are nothing but non-quantum analogues
of the well known quantum Heisenberg (''uncertainty'') relations. Also the
obtained fluctuations estimators support the idea that the Boltzmann's
constant $k$ has the signification of a generic indicator of stochasticity
for thermodynamic systems.

Pacs number(s): 05.20.-y, 05.40.-a, 05.70.-a, 41.20.Gz
\end{abstract}

\preprint{HEP/123-qed}

\narrowtext

\section{INTRODUCTION}

The literature from the last decades promoted some interesting attempts \cite
{1,2,3,4,5,6} in order \cite{4} ''to formulate a comprehensive and unified
theory of thermodynamic in the presence of fields''. As we know the
respective attempts approached the description of the thermodynamic systems
only in terms of macroscopic quantities regarded as deterministic variables,
unendowed with fluctuations.

On the other hand it is a well established fact that, due to their inner
microscopic structure, the thermodynamic systems must be characterized by
means of variables endowed with fluctuations. The mean values of the
respective variables coincide with the macroscopic quantities from the usual
thermodynamics. The alluded fluctuations require a description in terms of
some additional concepts of probabilistic nature (e.g. dispersions,
correlations, higher order moments). Such a description can be done in a
phenomenological or in a statistical-mechanics manner. In this paper we try
to develop a description of fluctuations specific for thermodynamic systems
in the presence of fields. Our description is done in a phenomenological
manner. We will appeal to usual procedures of phenomenological theory \cite
{2,4,7,8} as well as some ideas inspired by our earlier works on
fluctuations \cite{9,10}.

We start a search for a first approximation of the generalized distribution
of fluctuation probabilities. For this we use the concept of adequate
internal energy inspired by \cite{4} . A first application is focused on the
second order moments for the fluctuations in magnetizable continuum in a
magnetoquasistatic field. Afterward we will investigate briefly the
questions connected with the fluctuations in discrete systems in the
presence of magnetic field. By using the results of the alluded
investigations in a next section we will introduce a new class of
thermodynamic inequalities, in a manner similar with the one developed in 
\cite{9}.

\section{GENERAL THEORETICAL CONSIDERATIONS}

The phenomenological theory of fluctuations deals with real and continuous
variables endowed with an 'ad hoc' stochasticity (without any reference to
the microscopic structure of the thermodynamic systems). For small
fluctuations in the proximity of equilibrium states the corresponding
distribution of probabilities are given \cite{2,7,8} by the formula

\begin{equation}
w\sim \exp \left\{ \frac{\delta S^{\prime }}k\right\}  \label{eq:1}
\end{equation}

Here $\delta S^{\prime }=S^{\prime }(x)-S^{\prime }(\overline{x})$ denote
the variation of entropy due to the fluctuations, $x$ signify the set of the
macroscopic variables characterizing the system -with $\overline{x}$ as
equilibrium mean (or expected) value of $x$ and $k$ is the Boltzmann's
constant. The variation $\delta S^{\prime }$ refers to the ensemble of
thermodynamic system and its environment. It is given by

\begin{equation}
\delta S^{\prime }=-\frac{\delta W_{\min }}{T_{eq}}  \label{eq:2}
\end{equation}
with $T_{eq}=$equilibrium temperature and $\delta W_{\min }=$minimal work of
fluctuations.

{\it Observation}: In (\ref{eq:1}) as well as in all the subsequent formulas
for the probability distributions $w$ we are omiting the constants which
precedes the exponential functions. Evidently that, for all the cases, the
respective constants can be determined by imposing the normalization
condition for $w$.

Le us focus our attention on the systems in the presence of fields (as they
are viewed in \cite{4,5,6}. Then we introduce the work $\delta W_{\min }$
through the relation

\begin{equation}
\delta \widehat{U}=\sum_r\widehat{\xi _r}\,\,\delta Y_r+\sum_r\Psi
_l\,\delta Z_l+\delta W_{\min }  \label{eq:3}
\end{equation}

Here we used the following notations: $\delta A=$the variation (due to the
fluctuations) of the variable $A\,\,$; $\widehat{U}=$the internal energy in
the presence of fields; $\widehat{\xi _r}\,\,$ are the field dependent
intensive variables (e.g. $\widehat{\xi _1}=T=$temperature, while $-\widehat{%
\xi _2}=\widehat{p}$ and $\widehat{\xi _3}=\widehat{\zeta }$ denote the
field dependent pressure and chemical potential); $Y_r=$the usual extensive
thermodynamic variables (e.g. entropy $S$, volume $V$, number of particles $%
N $); $\Psi _l\,$and $Z_l$ denote the additional parameters due to the
presence of fields (e.g. $\Psi =V{\bf H}$, $Z={\bf B}$ with ${\bf H}$ and $%
{\bf B}$ strength respectively the induction of the magnetic field).

{\it Observation}: As above regarded the quantities $\Psi _l$ generally are
not intensive parameters conjugated with $Z_l$.

In (\ref{eq:3}) $\widehat{U}$ depends on $X_r$ and $Z_l$. Its total
differential is given by:

\begin{equation}
d\widehat{U}=\sum_r\widehat{\xi _r}\,\,dY_r+\sum_r\Psi _l\,dZ_l  \label{eq:4}
\end{equation}

For avoiding the irrelevant intricacy of the used formulas in the following
we introduce the compacted notation $\left\{ Y_r\right\} \cup \left\{
Z_l\right\} \stackrel{def}{=}\left\{ \eta _i\right\} $ and $\left\{ \widehat{%
\xi _r}\,\,\right\} \cup \left\{ \Psi _l\right\} \stackrel{def}{=}\left\{
\Phi i\right\} $. So from (\ref{eq:4}) one obtains:

\begin{equation}
\Phi _i=\left( \frac{\partial \widehat{U}}{\partial \eta _i}\right) _{eq}
\label{eq:5}
\end{equation}
where the index eq. denote the equilibrium value of the indexed quantity.

For the variation $\delta \widehat{U}$ in the second order approximation in
terms of the variations $\delta \eta _i$ one can write

\begin{equation}
\delta \widehat{U}=\sum_i\left( \frac{\partial \widehat{U}}{\partial \eta _i}%
\right) _{eq}\delta \eta _i+\frac 12\sum_{i,j}\left( \frac{\partial ^2%
\widehat{U}}{\partial \eta _i\,\partial \eta _j}\right) _{eq}\delta \eta
_i\,\delta \eta _j+...  \label{eq:6}
\end{equation}

By using the relations (\ref{eq:2}-\ref{eq:6}) from (\ref{eq:1}) one obtains:

\begin{equation}
w\sim \exp \left\{ -\frac 1{2kT}\sum_j\delta \Phi _j\,\delta \eta _j\right\}
\label{eq:7}
\end{equation}

For a given system, due to their physical nature, the variables from the
sets $\left\{ \Phi i\right\} $ and $\left\{ \eta _i\right\} $ are generally
interdependent. But for such a system always one can select a restrictive
set of physically independent variables $\left\{ X_a\right\} _{a=1}^n$. Then
(\ref{eq:7}) transcribes as the following multivariable Gaussian
distribution:

\begin{equation}
w\sim \exp \left\{ -\frac 12\sum_{a=1}^n\sum_{b=1}^n\alpha _{ab}\,\delta
X_a\,\delta X_b\right\}  \label{eq:8}
\end{equation}
where

\begin{equation}
\alpha _{ab}=\frac 1{kT}\sum_j\left( \frac{\partial \Phi _j}{\partial X_a}%
\right) _{eq}\left( \frac{\partial \eta _j}{\partial X_b}\right) _{eq}
\label{eq:9}
\end{equation}

From the previous relations it directly results the fact that $\alpha _{ab}$
are field dependent elements of a matrix $\alpha $.

It is the place to be noted the fact that in the Gaussian distribution (\ref
{eq:8}) the quantities $\left\{ X_a\right\} $ are considered as continuous
variables, each of them being defined in the range $\left( -\infty ,\infty
\right) $.

In the above mentioned Gaussian approximation the fluctuations of the
independent variables $\left\{ X_a\right\} $ are characterized by the
correlations

\begin{equation}
C_{ab}=\overline{\delta X_a\ \delta X_b}=\left( \alpha ^{-1}\right) _{ab}
\label{eq:10}
\end{equation}
where $\left( \alpha ^{-1}\right) _{ab}$ are the elements of the inverse of
the matrix $\alpha $ given by (\ref{eq:9}).

For any set of quantities $\left\{ Q_m\right\} $ expressible in terms of
independent variables $\left\{ X_a\right\} $ [i.e. $Q_m=Q_m\left( X_a\right) 
$] the fluctuations are characterized by the correlations

\begin{equation}
C_{ms}=\sum_a\sum_b\left( \frac{\partial Q_m}{\partial X_a}\right)
_{eq}\left( \frac{\partial Q_s}{\partial X_b}\right) _{eq}\left( \alpha
^{-1}\right) _{ab}  \label{eq:11}
\end{equation}

For $m=s$ the correlation $C_{mm}={\cal D}_m$ denote just the dispersion of
fluctuations for the quantity $Q_m$.

As in the case of fluctuations in the absence of fields \cite{9} the
correlations (\ref{eq:11}) constitute a real, symmetric and non-negative
definite matrix. Then from (\ref{eq:11}) one obtains the relation:

\begin{equation}
\det \left[ \sum_a\sum_b\left( \frac{\partial Q_m}{\partial X_a}\right)
_{eq}\left( \frac{\partial Q_s}{\partial X_b}\right) _{eq}\left( \alpha
^{-1}\right) _{ab}\right] \geq 0  \label{eq:12}
\end{equation}

Any concrete application of the above description of the fluctuations in the
presence of fields requires to consider the following steps:

\begin{itemize}
\item  to establish the adequate expression (\ref{eq:4}) for the total
differential $d\widehat{U}$ of the internal energy (for such a goal the
results from \cite{4,5,6} are highly recommendable).

\item  to evaluate the minimal work $\delta W_{\min }$ of fluctuations by
using (\ref{eq:3}).

\item  to identify the set of the independent variables $\left\{ X_a\right\} 
$.

\item  to take into account of the adequate equations of state and
thermodynamic Maxwell relations (in this sense as guides can be used the
works \cite{2,5,6,7,8} ).

\item  to compute effectively the adequate correlations $C_{ms}$ through the
relations (\ref{eq:10}) or (\ref{eq:11}).

\item  to find some relevant field dependent thermodynamic inequalities by
using (\ref{eq:12}).
\end{itemize}

In the Sections III and IV we will present some detailed exemplifications in
the above mentioned manner.

\section{EVALUATION OF SOME CORRELATIONS}

\subsection{The case of a magnetizable continuum in the presence of
magnetoquasistatic field}

In the alluded case we suppose that the magnetic energy is stored within the
frontiers of the system. The adequate expression of the internal energy
differential, as in \cite{4}, has the form

\begin{equation}
d\widehat{U}=TdS-\widehat{p}dV+\widehat{\zeta }dN+V{\bf H\,}d{\bf B}
\label{eq:13}
\end{equation}

In this relation, as well as in the subsequent ones, the implied symbols for
physical variables signify the mean values regarding an equilibrium state.
The quantities $\widehat{p}$ and $\widehat{\zeta }$ are intensive parameters
which are dependent on field. These parameters have \cite{4,5,6} various
expressions, depending on the approached situations.

The minimal work of the fluctuations is

\begin{equation}
\delta W_{\min }=\delta \widehat{U}-T\,\delta S+\widehat{p}\,\delta V-%
\widehat{\zeta }\,\delta N-V{\bf H\,}\delta {\bf B}  \label{eq:14}
\end{equation}

As independent variables we take $T,V,N$ and ${\bf B}$. Their independence
must be regarded in a thermodynamic sense, because they can be
interdependent from a statistical approach.

For the probability density (\ref{eq:8}) one obtains

\begin{equation}
w\sim \exp \left\{ -\frac 1{2kT}\left[ 
\begin{array}{c}
\left( \frac{\partial S}{\partial T}\right) _{V,N,{\bf B}}\left( \delta
T\right) ^2-\left( \frac{\partial \widehat{p}}{\partial V}\right) _{T,N,{\bf %
B}}\left( \delta V\right) ^2+\left( \frac{\partial \widehat{\zeta }}{%
\partial N}\right) _{T,V,{\bf B}}\left( \delta N\right) ^2+V\left( \frac{%
\partial H}{\partial B}\right) _{T,V,N}\left( \delta B\right) ^2 \\ 
+2\left( \frac{\partial \widehat{\zeta }}{\partial V}\right) _{T,N,{\bf B}%
}\delta V\delta N+2\left( \frac{\partial \left( VH\right) }{\partial V}%
\right) _{T,N,{\bf B}}\delta V\delta B+2V\left( \frac{\partial H}{\partial N}%
\right) _{T,V,{\bf B}}\delta N\delta B
\end{array}
\right] \right\}  \label{eq:15}
\end{equation}

The matrix of the correlation coefficients is

\begin{equation}
\left( \alpha \right) =\left( 
\begin{array}{llll}
\alpha _{11} & 0 & 0 & 0 \\ 
0 & \alpha _{22} & \alpha _{23} & \alpha _{24} \\ 
0 & \alpha _{32} & \alpha _{33} & \alpha _{34} \\ 
0 & \alpha _{42} & \alpha _{43} & \alpha _{44}
\end{array}
\right)  \label{eq:16}
\end{equation}

where

\begin{eqnarray*}
\alpha _{11} &=&\frac 1{kT}\left( \frac{\partial S}{\partial T}\right) _{V,N,%
{\bf B}};\quad \alpha _{22}=-\frac 1{kT}\left( \frac{\partial \widehat{p}}{%
\partial V}\right) _{T,N,{\bf B}} \\
\alpha _{33} &=&\frac 1{kT}\left( \frac{\partial \widehat{\zeta }}{\partial N%
}\right) _{T,V,{\bf B}};\quad \alpha _{44}=\frac V{kT}\left( \frac{\partial H%
}{\partial B}\right) _{T,V.N}=\frac V{kT\mu } \\
\alpha _{12} &=&\alpha _{21}=0;\quad \alpha _{13}=\alpha _{31}=0;\quad
\alpha _{14}=\alpha _{41}=0 \\
\alpha _{23} &=&\alpha _{32}=\frac 1{kT}\left( \frac{\partial \widehat{\zeta 
}}{\partial V}\right) _{T,N,{\bf B}} \\
\alpha _{24} &=&\alpha _{42}=\frac 1{kT}\left( \frac{\partial \left(
VH\right) }{\partial V}\right) _{T,N,{\bf B}}=\frac H{kT}\left[ 1+\frac \rho %
\mu \left( \frac{\partial \mu }{\partial \rho }\right) _T\right] \\
\alpha _{34} &=&\alpha _{43}=\frac V{kT}\left( \frac{\partial H}{\partial N}%
\right) _{T,V,{\bf B}}=-\frac H{kT\mu }\left( \frac{\partial \mu }{\partial
\rho }\right) _T
\end{eqnarray*}

In the above relations $\rho $ =denotes the particle number in the unity
volume $\left( \rho =\frac NV\right) $ and $\mu $ signifies the magnetic
permeability of the system.

Then for the dispersions and correlations one finds:

\begin{equation}
\overline{\left( \delta T\right) ^2}=\left( \alpha ^{-1}\right) _{11}=\frac 1%
{\alpha _{11}}  \label{eq:17}
\end{equation}

\begin{equation}
\overline{\left( \delta V\right) ^2}=\left( \alpha ^{-1}\right) _{22}=\frac{%
\left| 
\begin{array}{ll}
\alpha _{33} & \alpha _{34} \\ 
\alpha _{34} & \alpha _{44}
\end{array}
\right| }{\det \left| \beta \right| }  \label{eq:18}
\end{equation}

\begin{equation}
\overline{\left( \delta N\right) ^2}=\left( \alpha ^{-1}\right) _{33}=\frac{%
\left| 
\begin{array}{ll}
\alpha _{22} & \alpha _{24} \\ 
\alpha _{24} & \alpha _{44}
\end{array}
\right| }{\det \left| \beta \right| }  \label{eq:19}
\end{equation}

\begin{equation}
\overline{\left( \delta B\right) ^2}=\left( \alpha ^{-1}\right) _{44}=\frac{%
\left| 
\begin{array}{ll}
\alpha _{22} & \alpha _{23} \\ 
\alpha _{23} & \alpha _{33}
\end{array}
\right| }{\det \left| \beta \right| }  \label{eq:20}
\end{equation}

\begin{equation}
\overline{\delta T\,\delta V}=\overline{\delta T\,\delta N}=\overline{\delta
T\,\delta B}=0  \label{eq:21}
\end{equation}

\begin{equation}
\overline{\delta V\,\delta N}=\left( \alpha ^{-1}\right) _{23}=-\frac{\left| 
\begin{array}{ll}
\alpha _{23} & \alpha _{34} \\ 
\alpha _{24} & \alpha _{44}
\end{array}
\right| }{\det \left| \beta \right| }  \label{eq:22}
\end{equation}

\begin{equation}
\overline{\delta V\,\delta B}=\left( \alpha ^{-1}\right) _{24}=\frac{\left| 
\begin{array}{ll}
\alpha _{23} & \alpha _{33} \\ 
\alpha _{24} & \alpha _{34}
\end{array}
\right| }{\det \left| \beta \right| }  \label{eq:23}
\end{equation}

\begin{equation}
\overline{\delta N\,\delta B}=\left( \alpha ^{-1}\right) _{34}=-\frac{\left| 
\begin{array}{ll}
\alpha _{22} & \alpha _{23} \\ 
\alpha _{24} & \alpha _{34}
\end{array}
\right| }{\det \left| \beta \right| }  \label{eq:24}
\end{equation}

where

\begin{equation}
\det \left| \beta \right| =\left| 
\begin{array}{lll}
\alpha _{22} & \alpha _{23} & \alpha _{24} \\ 
\alpha _{23} & \alpha _{33} & \alpha _{34} \\ 
\alpha _{24} & \alpha _{34} & \alpha _{44}
\end{array}
\right|  \label{eq:25}
\end{equation}

Let us focus on some particular cases:

\subsubsection{$V=const,N=const.$}

In this case the matrix $\alpha $ is of the form

\begin{equation}
\left( \alpha \right) =\left( 
\begin{array}{ll}
\alpha _{11} & 0 \\ 
0 & \alpha _{22}
\end{array}
\right)  \label{eq:26}
\end{equation}

with

\[
\alpha _{11}=\frac 1{kT}\left( \frac{\partial S}{\partial T}\right) _{{\bf B}%
};\quad \alpha _{22}=\frac V{kT}\left( \frac{\partial H}{\partial B}\right)
_T 
\]

The entropy of the system in the presence of the magnetic field is given by 
\cite{1,3,5}

\begin{equation}
S=S_0+\frac 12VH^2\left( \frac{\partial \mu }{\partial T}\right) _\rho
\label{eq:27}
\end{equation}

where $S_0$ denote the entropy in the absence of the field.

Then, for the regarded case, for the dispersions and correlations of various
physical variables one obtains:

\begin{equation}
\overline{\left( \delta T\right) ^2}=\left( \alpha ^{-1}\right) _{11}=\frac{%
kT}{\left( \frac{\partial S}{\partial T}\right) _{{\bf B}}}=\frac{kT^2}{C_V+%
\frac 12TVH^2\left[ \left( \frac{\partial ^2\mu }{\partial T^2}\right) _\rho
-\frac 2\mu \left( \frac{\partial \mu }{\partial T}\right) _\rho ^2\right] }
\label{eq:28}
\end{equation}

\begin{equation}
\overline{\left( \delta B\right) ^2}=\left( \alpha ^{-1}\right) _{22}=\frac{%
kT}{V\left( \frac{\partial H}{\partial B}\right) _T}=\frac{kT\mu }V
\label{eq:29}
\end{equation}

\begin{equation}
\overline{\delta T\,\delta B}=0  \label{eq:30}
\end{equation}

\begin{equation}
\overline{\delta T\,\delta S}=\left( \frac{\partial S}{\partial T}\right) _{%
{\bf B}}\overline{\left( \delta T\right) ^2}=kT  \label{eq:31}
\end{equation}

\begin{eqnarray}
\overline{\delta T\,\delta H} &=&\left( \frac{\partial H}{\partial T}\right)
_{{\bf B}}\overline{\left( \delta T\right) ^2}=  \label{eq:32} \\
&=&-\frac H\mu \left( \frac{\partial \mu }{\partial T}\right) _\rho \frac{%
kT^2}{C_V+\frac 12TVH^2\left[ \left( \frac{\partial ^2\mu }{\partial T^2}%
\right) _\rho -\frac 2\mu \left( \frac{\partial \mu }{\partial T}\right)
_\rho ^2\right] }  \nonumber
\end{eqnarray}

\begin{equation}
\overline{\delta B\,\delta H}=\left( \frac{\partial H}{\partial B}\right) _T%
\overline{\left( \delta B\right) ^2}=\frac{kT}V  \label{eq:33}
\end{equation}

\begin{equation}
\overline{\delta S\,\delta B}=\left( \frac{\partial S}{\partial B}\right) _T%
\overline{\left( \delta B\right) ^2}=kTH\left( \frac{\partial \mu }{\partial
T}\right) _\rho  \label{eq:34}
\end{equation}

In the above formulas $C_V$ denotes the isochoric heat capacity: $%
C_V=T\left( \frac{\partial S_0}{\partial T}\right) _{V,\,N}$ . From (\ref
{eq:28}) it directly results that in the absence of field $\overline{\left(
\delta T\right) ^2}$ has the previously known expression \cite{7,8,9}.

\subsubsection{${\bf B}=const.$}

This case is associated with a constant magnetic flux. The quantities $T,V$
and $N$ will be regarded as random variables (i.e. endowed with
fluctuations).

In the considered case, according to \cite{4}, one can write

\begin{equation}
d\widehat{U}=TdS-\widehat{p}dV+\widehat{\zeta }dN  \label{eq:35}
\end{equation}

where

\begin{equation}
\widehat{p}=p_{{\bf B},N}=p-\frac 12{\bf HB-}\frac 12H^2\rho \left( \frac{%
\partial \mu }{\partial \rho }\right) _T  \label{eq:36}
\end{equation}

\begin{equation}
\widehat{\zeta }=\zeta _{{\bf B},V}=\zeta -\frac 12H^2\left( \frac{\partial
\mu }{\partial \rho }\right) _T  \label{eq:37}
\end{equation}

$p$ and $\zeta $ denote respectively the pressure and chemical potential in
the absence of field.

The matrix of the correlation coefficients has the form

\begin{equation}
\left( \alpha \right) =\left( 
\begin{array}{lll}
\alpha _{11} & 0 & 0 \\ 
0 & \alpha _{22} & \alpha _{23} \\ 
0 & \alpha _{23} & \alpha _{33}
\end{array}
\right)  \label{eq:38}
\end{equation}

By using (\ref{eq:11}) through of some uncomplicated algebraic operations
one finds

\begin{equation}
\overline{\left( \delta T\right) ^2}=\left( \alpha ^{-1}\right) _{11}=\frac{%
kT}{\left( \frac{\partial S}{\partial T}\right) _{V,N,{\bf B}}}
\label{eq:39}
\end{equation}

\begin{equation}
\overline{\left( \delta V\right) ^2}=\left( \alpha ^{-1}\right) _{22}=\frac{%
\alpha _{33}}{\alpha _{22}\alpha _{33}-\alpha _{23}^2}=-kT\frac{\left( \frac{%
\partial \widehat{\zeta }}{\partial N}\right) _{T,V,{\bf B}}}{\left( \frac{%
\partial \widehat{p}}{\partial V}\right) _{T,N,{\bf B}}\left( \frac{\partial 
\widehat{\zeta }}{\partial N}\right) _{T,V,{\bf B}}+\left( \frac{\partial 
\widehat{\zeta }}{\partial V}\right) _{T,N,{\bf B}}^2}  \label{eq:40}
\end{equation}

\begin{equation}
\overline{\left( \delta N\right) ^2}=\left( \alpha ^{-1}\right) _{33}=\frac{%
\alpha _{22}}{\alpha _{22}\alpha _{33}-\alpha _{23}^2}=kT\frac{\left( \frac{%
\partial \widehat{p}}{\partial V}\right) _{T,N,{\bf B}}}{\left( \frac{%
\partial \widehat{p}}{\partial V}\right) _{T,N,{\bf B}}\left( \frac{\partial 
\widehat{\zeta }}{\partial N}\right) _{T,V,{\bf B}}+\left( \frac{\partial 
\widehat{\zeta }}{\partial V}\right) _{T,N,{\bf B}}^2}  \label{eq:41}
\end{equation}

\begin{equation}
\overline{\delta V\,\delta N}=\left( \alpha ^{-1}\right) _{23}=\frac{\alpha
_{23}}{\alpha _{23}^2-\alpha _{22}\alpha _{33}}=kT\frac{\left( \frac{%
\partial \widehat{\zeta }}{\partial V}\right) _{T,N,{\bf B}}}{\left( \frac{%
\partial \widehat{\zeta }}{\partial V}\right) _{T,N,{\bf B}}^2+\left( \frac{%
\partial \widehat{p}}{\partial V}\right) _{T,N,{\bf B}}\left( \frac{\partial 
\widehat{\zeta }}{\partial N}\right) _{T,V,{\bf B}}}  \label{eq:42}
\end{equation}

\begin{equation}
\overline{\delta T\,\delta V}=0;\overline{\delta T\,\delta N}=0
\label{eq:43}
\end{equation}

The formulas (39)-(42) imply the following relations:

\begin{equation}
\left( \frac{\partial S}{\partial T}\right) _{V,N,{\bf B}}=\frac{C_V}T+\frac %
12VH^2\left[ \left( \frac{\partial ^2\mu }{\partial T^2}\right) _\rho -\frac %
2\mu \left( \frac{\partial \mu }{\partial T}\right) _\rho ^2\right]
\label{eq:44}
\end{equation}

\begin{equation}
\left( \frac{\partial \widehat{\zeta }}{\partial N}\right) _{T,V,{\bf B}%
}=\left( \frac{\partial \zeta }N\right) _{T,V}+\frac{H^2}{\mu V}\left( \frac{%
\partial \mu }{\partial \rho }\right) _T^2-\frac 12\frac{H^2}V\left( \frac{%
\partial ^2\mu }{\partial \rho ^2}\right) _T  \label{eq:45}
\end{equation}

\begin{equation}
\left( \frac{\partial \widehat{p}}V\right) _{T,N,{\bf B}}=\left( \frac{%
\partial p}{\partial V}\right) _{T,N}-\frac{H^2\rho ^2}{\mu V}\left( \frac{%
\partial \mu }{\partial \rho }\right) _T^2+\frac 12\frac{H^2\rho ^2}V\left( 
\frac{\partial ^2\mu }{\partial \rho ^2}\right) _T  \label{eq:46}
\end{equation}

\begin{equation}
\left( \frac{\partial \widehat{\zeta }}{\partial V}\right) _{T,N,{\bf B}%
}=\left( \frac{\partial \zeta }{\partial N}\right) _{T,N}-\frac{H^2\rho }{%
\mu V}\left( \frac{\partial \mu }{\partial \rho }\right) _T^2+\frac 12\frac{%
H^2\rho }V\left( \frac{\partial ^2\mu }{\partial \rho ^2}\right) _T
\label{eq:47}
\end{equation}

\subsubsection{${\bf H}=const.$}

In this case the courant densities (the sources of the magnetic field) are
constant.. We search the parameters of the fluctuations for the quantities $%
T,V$ and $N$.

The differential of the internal energy is given by (\ref{eq:35}) where
according to \cite{4}

\begin{equation}
\widehat{p}=p_{{\bf H},N}=p-\frac 12{\bf HB}+\frac 12H^2\rho \left( \frac{%
\partial \mu }{\partial \rho }\right) _T  \label{eq:48}
\end{equation}

\begin{equation}
\widehat{\zeta }=\zeta _{{\bf H},V}=\zeta +\frac 12H^2\left( \frac{\partial
\mu }{\partial \rho }\right) _T  \label{eq:49}
\end{equation}

In the approached case, as it was proved in \cite{6}, the entropy is

\begin{equation}
S\left( {\bf H}=const.\right) =S_0-\frac 12VH^2\left( \frac{\partial \mu }{%
\partial T}\right) _\rho  \label{eq:50}
\end{equation}

With such an expression for entropy one obtains

\begin{equation}
\overline{\left( \delta T\right) ^2}=\frac{kT}{\left( \frac{\partial S}{%
\partial T}\right) _{V,N,{\bf H}}}=\frac{kT^2}{C_V-\frac 12TVH^2\left( \frac{%
\partial ^2\mu }{\partial T^2}\right) _\rho }  \label{eq:51}
\end{equation}

\begin{equation}
\overline{\left( \delta V\right) ^2}=-kT\frac{\left( \frac{\partial \widehat{%
\zeta }}{\partial N}\right) _{T,V,{\bf H}}}{\left( \frac{\partial \widehat{p}%
}{\partial V}\right) _{T,N,{\bf H}}\left( \frac{\partial \widehat{\zeta }}{%
\partial N}\right) _{T,V,{\bf H}}+\left( \frac{\partial \widehat{\zeta }}{%
\partial V}\right) _{T,N,{\bf H}}^2}  \label{eq:52}
\end{equation}

\begin{equation}
\overline{\left( \delta N\right) ^2}=kT\frac{\left( \frac{\partial \widehat{p%
}}{\partial V}\right) _{T,N,{\bf H}}}{\left( \frac{\partial \widehat{p}}{%
\partial V}\right) _{T,N,{\bf H}}\left( \frac{\partial \widehat{\zeta }}{%
\partial N}\right) _{T,V,{\bf H}}+\left( \frac{\partial \widehat{\zeta }}{%
\partial V}\right) _{T,N,{\bf H}}^2}  \label{eq:53}
\end{equation}

\begin{equation}
\overline{\delta V\,\delta N}=kT\frac{\left( \frac{\partial \widehat{\zeta }%
}{\partial V}\right) _{T,N,{\bf H}}}{\left( \frac{\partial \widehat{\zeta }}{%
\partial V}\right) _{T,N,{\bf H}}^2+\left( \frac{\partial \widehat{p}}{%
\partial V}\right) _{T,N,{\bf H}}\left( \frac{\partial \widehat{\zeta }}{%
\partial N}\right) _{T,V,{\bf H}}}  \label{eq:54}
\end{equation}

where

\quad 
\begin{equation}
\left( \frac{\partial \widehat{\zeta }}{\partial N}\right) _{T,V,{\bf H}%
}=\left( \frac{\partial \zeta }{\partial N}\right) _{T,V}+\frac 12\frac{H^2}V%
\left( \frac{\partial ^2\mu }{\partial \rho ^2}\right) _T  \label{eq:55}
\end{equation}

\begin{equation}
\left( \frac{\partial \widehat{p}}{\partial V}\right) _{T,N,{\bf H}}=\left( 
\frac{\partial p}{\partial V}\right) _{T,N}-\frac 12\frac{H^2\rho ^2}V\left( 
\frac{\partial ^2\mu }{\partial \rho ^2}\right) _T  \label{eq:56}
\end{equation}

\begin{equation}
\left( \frac{\partial \widehat{\zeta }}{\partial V}\right) _{T,N,{\bf H}%
}=\left( \frac{\partial \zeta }{\partial V}\right) _{T,N}-\frac 12\frac{%
H^2\rho }V\left( \frac{\partial ^2\mu }{\partial \rho ^2}\right) _T
\label{eq:57}
\end{equation}

This subsection must be completed with the following important
specifications:

\begin{itemize}
\item  All the above relations refer to the case of linear magnetic
materials, for which $\mu $ is independent of $H$ and depends only on the
temperature $T$ and particle density $\rho $.

\item  For practical purposes it is more useful that the fluctuation to be
evaluated for magnetization {\bf M} but not for magnetic induction. Such a
choice of evaluation can be done if one substract from the generalized
internal energy the excitation energy corresponding to the magnetic field in
vacuo, i.e. by introducing the function

\begin{equation}
\widehat{U}^{*}=\widehat{U}-\frac 12V\mu _0H^2  \label{eq:58}
\end{equation}

The alluded choice is possible only if it is neglected the change of $H$ due
to the presence of the magnetic system. Rigorously it must to be substract
from $\widehat{U}$ the quantity $\frac 12V\mu _0{\cal H}^2$ , where ${\cal H}
$ is the field strength generated in the vacuo by the sources , in absence
of the system (for similar considerations see \cite{1}). For most of the
systems the magnetic susceptibility has a low value. Therefore one can be
neglected the deformation of the field generated by the system presence. In
such case one obtains:

\begin{equation}
d\widehat{U}^{*}=TdS-\widehat{p}dV+\widehat{\zeta }dN+V{\bf H}d\left( \mu _0%
{\bf M}\right)   \label{eq:59}
\end{equation}

where

\begin{equation}
{\bf M}=\chi _m{\bf H}  \label{eq:60}
\end{equation}

\begin{equation}
\widehat{p}=p-\frac 12\mu _0{\bf HM}-\frac 12H^2\rho \mu _0\frac{\partial
\chi _m}{\partial \rho }  \label{eq:61}
\end{equation}

\begin{equation}
\widehat{\zeta }=\zeta -\frac 12H^2\mu _0\frac{\partial \chi _m}{\partial
\rho }  \label{eq:62}
\end{equation}

In the last three relations $\mu _0$ and $\chi _m$ denote the magnetic
permeability of vacuo and respectively the magnetic susceptibility.

When $V$ and $N$ have constant values one finds the restricted formula:

\begin{equation}
\overline{\left( \delta M\right) ^2}=\frac{kT\mu _0\chi _m}V  \label{eq:63}
\end{equation}

known also from \cite{2}.

\item  In the case of a dielectric continuum the expressions of the
fluctuations for various physical quantities can be obtained from the from
the above presented formulas by means of the following substitutions:${\bf B}%
\longrightarrow {\bf D};{\bf H}\longrightarrow {\bf E}$; $\mu
\longrightarrow \varepsilon ,$ where ${\bf D}$ is the electric induction, $%
{\bf E}$ is the electric field strength and $\varepsilon $ the dielectric
permittivity. Also it must to be replaced $\mu _0{\bf M}$ with ${\bf P}$
(dielectric polarization) and $\chi _m$ with $\chi _e$ (dielectric
susceptibility).

\item  It is easy to observe that the dispersions $\overline{\left( \delta
A\right) ^2}$ and correlations $\overline{\delta A\,\delta B}$ from the
above established formulas appears as products of Boltzmann's constant $k$
with expressions which contains exclusively mean values of the random
variables. To be noted that the respective values identify themselves with
the variables from the ordinary (non-stochastic) thermodynamics.
\end{itemize}

\subsection{The case of discrete systems}

The discrete systems are characterized by the fact that the field lines
extend also in the outside of the system. Then the field energy is stored
both in the inside and outside of the system. For such systems additionally
to the deterministic- thermodynamic approach their study must be completed
also with an investigation of the fluctuations\ (i.e. an evaluation of their
stochastic characteristics). In the following we will approach the
respective investigation for a particular case investigated determinist by
Y. Zimmels \cite{5}. The respective case regards a sphere of radius $R$
placed in an external uniform magnetic field of strength ${\bf H}_0$.

For generalized internal energy of the sphere we have

\begin{equation}
dU_1=TdS_1-\widehat{p}dV_1+\widehat{\zeta }dN_1+\psi dB_1  \label{eq:64}
\end{equation}

This formula imply the following relations, taken by \cite{5}:

\begin{equation}
\psi =V_1\frac{B_1}{\mu _s^{\prime }}=V_1H_1\sqrt{\frac{\mu _s}{\mu
_s^{\prime }}}  \label{eq:65}
\end{equation}

\begin{equation}
\mu _s=\frac{\mu _1}9\left( \frac{\mu _1}{\mu _2}-2\frac{\mu _2}{\mu _1}%
+1\right)  \label{eq:66}
\end{equation}

\begin{equation}
\frac 1{\mu _s^{\prime }}=\frac 19\left( \frac 1{\mu _2}+\frac 1{\mu _1}-2%
\frac{\mu _2}{\mu _1^2}\right)  \label{eq:67}
\end{equation}

In the above relations as well as in the following ones the indexes 1 and 2
refer to the system (sphere) respectively to the environment.

Let us now discuss some particular situations:

\subsubsection{$V_1=const.,N_1=const$}

In this situation the entropy of the sphere, as in \cite{5}, is

\begin{equation}
S_1=S_{01}+\frac 1{18}V_1B_1^2\left[ \alpha \frac{\partial \mu _2}{\partial T%
}-\beta \frac{\partial \mu _1}{\partial T}\right]  \label{eq:68}
\end{equation}

where $S_{01}$ denotes the entropy in the absence of the field and

\begin{equation}
\alpha =\frac 1{\mu _2^2}+\frac 2{\mu _1^2}\quad ;\qquad \beta =\frac{4\mu _2%
}{\mu _1^3}-\frac 1{\mu _1^2}  \label{eq:69}
\end{equation}

The relation (\ref{eq:65}-\ref{eq:69}) referee to the discrete system in an
equilibrium thermodynamic state.

If one takes as independent variables the quantities $T$ and $B_1$, for
their fluctuations one obtains:

\[
\overline{\left( \delta T\right) ^2}=\frac{kT}{\left( \frac{\partial S_1}{%
\partial T}\right) _{{\bf B}_1}}= 
\]

\begin{equation}
=kT^2\left\{ C_V+\frac 1{18}TV_1B_1^2\left[ \alpha \frac{\partial ^2\mu _2}{%
\partial T^2}-\beta \frac{\partial ^2\mu _1}{\partial T^2}+\frac 2{\mu _1^3}%
\left( \frac{\partial \mu _1}{\partial T}\right) ^2\left( \frac{6\mu _2}{\mu
_1}-1\right) -\frac 2{\mu _2^3}\left( \frac{\partial \mu _2}{\partial T}%
\right) ^2-\frac 8{\mu _1^3}\frac{\partial \mu _1}{\partial T}\frac{\partial
\mu _2}{\partial T}\right] \right\} ^{-1}  \label{eq:70}
\end{equation}

\begin{equation}
\overline{\left( \delta B_1\right) ^2}=\frac{kT\mu _s^{\prime }}{V_1}=\frac{%
kT}{V_1}\left\{ \frac 19\left( \frac 1{\mu _2}+\frac 1{\mu _1}-2\frac{\mu _2%
}{\mu _1^2}\right) \right\} ^{-1}  \label{eq:71}
\end{equation}

These relations show that for the discrete systems the fluctuations of the
macroscopic parameters depend on the permeabilities of both the system and
the environment.

\subsubsection{${\bf B}_1=const.$}

In this situation as in \cite{5} we take

\begin{equation}
dU_1=TdS_1-\widehat{p}dV_1+\widehat{\zeta }dN_1  \label{eq:72}
\end{equation}

with

\begin{equation}
\widehat{p}=p_{{\bf B}_1,\,N}=p-\frac{B_1^2}{2\mu _s^{\prime }}+\frac{B_1^2}{%
18}\left[ \alpha \rho _2\frac{V_1}{V_2}\frac{\partial \mu _2}{\partial \rho
_2}+\beta \rho _1\frac{\partial \mu _1}{\partial \rho _1}\right]
\label{eq:73}
\end{equation}

and

\begin{equation}
\widehat{\zeta }=\zeta _{{\bf B}_1,V}=\zeta +\frac{B_1^2}{18}\left[ \frac{V_1%
}{V_2}\alpha \frac{\partial \mu _2}{\partial \rho _2}+\beta \frac{\partial
\mu _1}{\partial \rho _1}\right]  \label{eq:74}
\end{equation}

It results that for$\overline{\left( \delta T\right) ^2}$ one obtains the
same expression as in previous situation. This because the entropy is given
also by (\ref{eq:68}).

The dispersions of $V_1$ and $N_1$ are:

\begin{equation}
\overline{\left( \delta V_1\right) ^2}=-kT\frac{\left( \frac{\partial 
\widehat{\zeta }}{\partial N_1}\right) _{T,V_1,{\bf B}_1}}{\left( \frac{%
\partial \widehat{\zeta }}{\partial N_1}\right) _{T,V_1,{\bf B}_1}\left( 
\frac{\partial \widehat{p}}{\partial V_1}\right) _{T,N_1,{\bf B}_1}+\left( 
\frac{\partial \widehat{\zeta }}{\partial V_1}\right) _{T,N_1,{\bf B}_1}^2}
\label{eq:75}
\end{equation}

\begin{equation}
\overline{\left( \delta N_1\right) ^2}=kT\frac{\left( \frac{\partial 
\widehat{p}}{\partial V_1}\right) _{T,N_1,{\bf B}_1}}{\left( \frac{\partial 
\widehat{p}}{\partial V_1}\right) _{T,N_1,{\bf B}_1}\left( \frac{\partial 
\widehat{\zeta }}{\partial N_1}\right) _{T,V_1,{\bf B}_1}+\left( \frac{%
\partial \widehat{\zeta }}{\partial V_1}\right) _{T,N_1,{\bf B}_1}^2}
\label{eq:76}
\end{equation}

These expressions are only in a formal analogy with the corresponding ones
for magnetizable continuum, because they imply specific particularities
through the following partial derivatives.

\begin{equation}
\left( \frac{\partial \widehat{p}}{\partial V_1}\right) _{T,N_1,{\bf B}%
_1}=\left( \frac{\partial p}{\partial V_1}\right) _{T,N_1}+\frac{B_1^2}{18}%
\left[ \frac{\alpha \rho _2}{V_2}\frac{\partial \mu _2}{\partial \rho _2}+%
\frac{\beta \rho _1}{V_1}\frac{\partial \mu _1}{\partial \rho _1}+\frac %
\partial {\partial V_1}\left( \alpha \rho _2\frac{V_1}{V_2}\frac{\partial
\mu _2}{\partial \rho _2}+\beta \rho _1\frac{\partial \mu _1}{\partial \rho
_1}\right) \right]  \label{eq:77}
\end{equation}

\begin{equation}
\left( \frac{\partial \widehat{\zeta }}{\partial N_1}\right) _{T,V_1,{\bf B}%
_1}=\left( \frac{\partial \zeta }{\partial N_1}\right) _{T,V_1}+\frac{B_1^2}{%
18}\frac \partial {\partial N_1}\left( \frac{V_1}{V_2}\alpha \frac{\partial
\mu _2}{\partial \rho _2}+\beta \frac{\partial \mu _1}{\partial \rho _1}%
\right)  \label{eq:78}
\end{equation}

\begin{equation}
\left( \frac{\partial \widehat{\zeta }}{\partial V_1}\right) _{T,N_1,{\bf B}%
_1}=\left( \frac{\partial \zeta }{\partial V_1}\right) _{T,N_1}+\frac{B_1^2}{%
18}\frac \partial {\partial V_1}\left( \frac{V_1}{V_2}\alpha \frac{\partial
\mu _2}{\partial \rho _2}+\beta \frac{\partial \mu _1}{\partial \rho _1}%
\right)  \label{eq:79}
\end{equation}

\subsubsection{${\bf H}_1=const.$}

Now for the used quantities as in \cite{5} we have the expressions:

\begin{equation}
\widehat{p}=p_{{\bf H},N}=p-\frac 12H_1^2\mu _s+\frac{H_1^2}{18}\left( \beta
^{\prime }\rho _1\frac{\partial \mu _1}{\partial \rho _1}+\alpha ^{\prime
}\rho _2\frac{V_1}{V_2}\frac{\partial \mu _2}{\partial \rho _2}\right)
\label{eq:80}
\end{equation}

\begin{equation}
\widehat{\zeta }=\zeta _{{\bf H},V}=\zeta +\frac{H_1^2}{18}\left( \beta
^{\prime }\frac{\partial \mu _1}{\partial \rho _1}+\alpha ^{\prime }\frac{V_1%
}{V_2}\frac{\partial \mu _2}{\partial \rho _2}\right)  \label{eq:81}
\end{equation}

\begin{equation}
\alpha ^{\prime }=\frac{\mu _1^2}{\mu _2^2}+2\quad ;\quad \beta ^{\prime }=%
\frac{2\mu _1}{\mu _2}+1  \label{eq:82}
\end{equation}

\begin{equation}
S_1\left( {\bf H}_1=const.\right) =S_{01}+\frac 1{18}V_1H_1^2\left( -\beta
^{\prime }\frac{\partial \mu _1}{\partial T}+\alpha ^{\prime }\frac{\partial
\mu _2}{\partial T}\right)  \label{eq:83}
\end{equation}

Then for the fluctuation of the temperature one obtains:

\[
\overline{\left( \delta T\right) ^2}=\frac{kT}{\left( \frac{\partial S}{%
\partial T}\right) _{V_1,\,N_1,\,{\bf H}_1}}= 
\]

\begin{equation}
=kT^2\left\{ C_V+\frac 1{18}TV_1H_1^2\left[ \alpha ^{\prime }\frac{\partial
^2\mu _2}{\partial T^2}-\beta ^{\prime }\frac{\partial ^2\mu _1}{\partial T^2%
}-\frac 2{\mu _2}\left( \frac{\partial \mu _1}{\partial T}\right) ^2-\frac{%
2\mu _1^2}{\mu _2^3}\left( \frac{\partial \mu _2}{\partial T}\right) ^2+%
\frac{4\mu _1}{\mu _2^2}\frac{\partial \mu _1}{\partial T}\frac{\partial \mu
_2}{\partial T}\right] \right\} ^{-1}  \label{eq:84}
\end{equation}

Correspondingly for the fluctuation of $V_1$ and $N_1$ we find:

\begin{equation}
\overline{\left( \delta V_1\right) ^2}=-kT\frac{\left( \frac{\partial 
\widehat{\zeta }}{\partial N_1}\right) _{T,V_1,{\bf H}_1}}{\left( \frac{%
\partial \widehat{\zeta }}{\partial N_1}\right) _{T,V_1,{\bf H}_1}\left( 
\frac{\partial \widehat{p}}{\partial V_1}\right) _{T,N_1,{\bf H}_1}+\left( 
\frac{\partial \widehat{\zeta }}{\partial V_1}\right) _{T,N_1,{\bf H}_1}^2}
\label{eq:85}
\end{equation}

\begin{equation}
\overline{\left( \delta N_1\right) ^2}=kT\frac{\left( \frac{\partial 
\widehat{p}}{\partial V_1}\right) _{T,N_1,{\bf H}_1}}{\left( \frac{\partial 
\widehat{p}}{\partial V_1}\right) _{T,N_1,{\bf H}_1}\left( \frac{\partial 
\widehat{\zeta }}{\partial N_1}\right) _{T,V_1,{\bf H}_1}+\left( \frac{%
\partial \widehat{\zeta }}{\partial V_1}\right) _{T,N_1,{\bf H}_1}^2}
\label{eq:86}
\end{equation}

where

\begin{equation}
\left( \frac{\partial \widehat{p}}{\partial V_1}\right) _{T,\,N_1,\,{\bf H}%
_1}=\left( \frac{\partial p}{\partial V_1}\right) _{T,\,N_1}+\frac{H_1^2}{18}%
\left[ \frac{\alpha ^{\prime }\rho _2}{V_2}\frac{\partial \mu _2}{\partial
\rho _2}+\beta ^{\prime }\frac{\rho _1}{V_1}\frac{\partial \mu _1}{\partial
\rho _1}+\frac \partial {\partial V_1}\left( \beta ^{\prime }\rho _1\frac{%
\partial \mu _1}{\partial \rho _1}+\alpha ^{\prime }\rho _2\frac{V_1}{V_2}%
\frac{\partial \mu _2}{\partial \rho _2}\right) \right]  \label{eq:87}
\end{equation}

\begin{equation}
\left( \frac{\partial \widehat{\zeta }}{\partial N_1}\right) _{T,\,V_1,\,%
{\bf H}_1}=\left( \frac{\partial \zeta }{\partial N_1}\right) _{T,\,V_1}+%
\frac{H_1^2}{18}\frac \partial {\partial N_1}\left( \beta ^{\prime }\frac{%
\partial \mu _1}{\partial \rho _1}+\alpha ^{\prime }\frac{V_1}{V_2}\frac{%
\partial \mu _2}{\partial \rho _2}\right)  \label{eq:88}
\end{equation}

\begin{equation}
\left( \frac{\partial \widehat{\zeta }}{\partial V_1}\right) _{T,\,N_1,\,%
{\bf H}_1}=\left( \frac{\partial \zeta }{\partial V_1}\right) _{T,\,N_1}+%
\frac{H_1^2}{18}\frac \partial {\partial V_1}\left( \beta ^{\prime }\frac{%
\partial \mu _1}{\partial \rho _1}+\alpha ^{\prime }\frac{V_1}{V_2}\frac{%
\partial \mu _2}{\partial \rho _2}\right)  \label{eq:89}
\end{equation}

It is interesting at this point to discuss the extreme case of infinite
permeability for the sphere, when $\frac{\mu _2}{\mu _1}\rightarrow 0$. Then
the field energy associated with the discrete system is stored exclusively
in the outside of the system. In the alluded case

\begin{equation}
\alpha =\frac 1{\mu _2^2}\;;\quad \beta =0\;;\quad H_1=0\;;\quad \lim_{\mu
_1\rightarrow \infty }B_1=3\mu _2H_0\;;\quad \mu _s^{\prime }=9\mu _2
\label{eq:90}
\end{equation}

If $V_1$ and $N_1$ are constant for the temperature and the magnetic
induction one obtains:

\begin{equation}
\overline{\left( \delta T\right) ^2}=kT^2\left\{ C_V+\frac 12TV_1H_0^2\left[ 
\frac{\partial ^2\mu _2}{\partial T^2}-\frac 2{\mu _2}\left( \frac{\partial
\mu _2}{\partial T}\right) ^2\right] \right\} ^{-1}  \label{eq:91}
\end{equation}

\begin{equation}
\overline{\left( \delta B_1\right) ^2}=\frac{9kT\mu _2}{V_1}  \label{eq:92}
\end{equation}

For ${\bf B}_1=const.$, if one consider $V_1\ll V_2$ the parameters $%
\widehat{p}$ and $\widehat{\zeta }$ have the following expressions

\begin{equation}
\widehat{p}=p-\frac{\mu _2H_0^2}2=p-\frac 1{18}\frac{B_1^2}{\mu _2}
\label{eq:93}
\end{equation}

\begin{equation}
\widehat{\zeta }=\zeta  \label{eq:94}
\end{equation}

given in \cite{5}.

The fluctuations of $V_1$ and $N_1$ are characterized by the dispersions:

\begin{equation}
\overline{\left( \delta V_1\right) ^2}=-kT\frac{\left( \frac{\partial \zeta 
}{\partial N_1}\right) _{T,\,V_1}}{\left( \frac{\partial \zeta }{\partial N_1%
}\right) _{T,\,V_1}\left[ \left( \frac{\partial p}{\partial V_1}\right)
_{T,\,N_1}+\frac 12H_0^2\frac{\rho _2}{V_2}\frac{\partial \mu _2}{\partial
\rho _2}\right] +\left( \frac{\partial \zeta }{\partial V_1}\right)
_{T,\,N_1}^2}  \label{eq:95}
\end{equation}

\begin{equation}
\overline{\left( \delta N_1\right) ^2}=kT\frac{\left( \frac{\partial p}{%
\partial V_1}\right) _{T,\,N_1}+\frac 12H_0^2\frac{\rho _2}{V_2}\frac{%
\partial \mu _2}{\partial \rho _2}}{\left( \frac{\partial \zeta }{\partial
N_1}\right) _{T,\,V_1}\left[ \left( \frac{\partial p}{\partial V_1}\right)
_{T,\,N_1}+\frac 12H_0^2\frac{\rho _2}{V_2}\frac{\partial \mu _2}{\partial
\rho _2}\right] +\left( \frac{\partial \zeta }{\partial V_1}\right)
_{T,\,N_1}^2}  \label{eq:96}
\end{equation}

In the same extreme case for ${\bf H}_1=const.$ we obtain:

\begin{equation}
\widehat{p}=p-\frac 12\mu _2H_0^2  \label{eq:97}
\end{equation}

\begin{equation}
\widehat{\zeta }=\zeta  \label{eq:98}
\end{equation}

\begin{equation}
\overline{\left( \delta V_1\right) ^2}=-kT\frac{\left( \frac{\partial \zeta 
}{\partial N_1}\right) _{T,\,V_1}}{\left( \frac{\partial \zeta }{\partial N_1%
}\right) _{T,\,V_1}\left[ \left( \frac{\partial p}{\partial V_1}\right)
_{T,\,N_1}-\frac 12H_0^2\frac{\rho _2}{V_2}\frac{\partial \mu _2}{\partial
\rho _2}\right] +\left( \frac{\partial \zeta }{\partial V_1}\right)
_{T,\,N_1}^2}  \label{eq:99}
\end{equation}

\begin{equation}
\overline{\left( \delta N_1\right) ^2}=kT\frac{\left( \frac{\partial p}{%
\partial V_1}\right) _{T,\,N_1}-\frac 12H_0^2\frac{\rho _2}{V_2}\frac{%
\partial \mu _2}{\partial \rho _2}}{\left( \frac{\partial \zeta }{\partial
N_1}\right) _{T,\,V_1}\left[ \left( \frac{\partial p}{\partial V_1}\right)
_{T,\,N_1}-\frac 12H_0^2\frac{\rho _2}{V_2}\frac{\partial \mu _2}{\partial
\rho _2}\right] +\left( \frac{\partial \zeta }{\partial V_1}\right)
_{T,\,N_1}^2}  \label{eq:100}
\end{equation}

It must specified that in the alluded case ( $\frac{\mu _2}{\mu _1}%
\rightarrow 0$) the fixed value of $H_1$ means $H_1=0$.

This subsection must be completed with the following specifications:

\begin{itemize}
\item  The most important fact is that, in the case of discrete systems, the
fluctuations of the intrinsic parameters of the systems depends on the
magnetic permeability of the environment.

\item  For dielectric systems the evaluation of the fluctuations can be
obtained by similar considerations, by means of the substitutions: ${\bf H}%
\rightarrow {\bf E},\,{\bf B}\rightarrow {\bf D},\,\mu \rightarrow
\varepsilon $.

\item  As regard the here discussed dispersions $\overline{\left( \delta
A\right) ^2}$ and the correlations $\overline{\delta A\,\delta B}$ we have
to note the same specification as the last one from the previous subsection.
\end{itemize}

\section{THERMODYNAMIC INEQUALITIES FOR SYSTEMS IN THE PRESENCE OF FIELDS}

As it is known \cite{9}, the correlation coefficient constitute the elements
of a non-negatively defined matrix. This fact is expressed by the
inequalities

\begin{equation}
\det \left| C_{ab}\right| >0  \label{eq:101}
\end{equation}

\begin{equation}
\det \left| C_{ab}^{-1}\right| >0  \label{eq:102}
\end{equation}

where $C_{ab}^{-1}$ denote the inverse of the matrix $C_{ab}$.

By using (\ref{eq:101}) and (\ref{eq:102}) it is possible to obtain a lot of
thermodynamic inequalities. In order to exemplify the respective
possibility, in the Table \ref{table1} we included some such inequalities
which refer to a magnetizable continuum situated in a magnetoquasistatic
field.

If one considers separately only two variables, $X_1$ and $X_2$, the formula
(\ref{eq:101}) gives

\begin{equation}
\overline{\left( \delta X_1\right) ^2}\;\overline{\left( \delta X_2\right) ^2%
}>\left( \overline{\delta X_1\delta X_2}\right) ^2  \label{eq:103}
\end{equation}

This kind of relations, in our opinion \cite{9,10,11,12,13,14}, are
completely similar with the well known Heisenberg's (''uncertainty'')
relations from quantum mechanics.

In the end of this section we illustrate the relation (\ref{eq:103}) for
some concrete cases.

For a magnetizable continuum in a magnetoquasistatic field, when $N$ and $V$
are constant, from (\ref{eq:103}) we find the following inequalities:

\begin{equation}
\overline{\left( \delta T\right) ^2}\;\overline{\left( \delta B\right) ^2}>0
\label{eq:104}
\end{equation}

\begin{equation}
\overline{\left( \delta T\right) ^2}\;\overline{\,\left( \delta S\right) ^2}%
>k^2T^2  \label{eq:105}
\end{equation}

\begin{equation}
\,\overline{\left( \delta B\right) ^2}\;\overline{\left( \delta H\right) ^2}>%
\frac{k^2T^2}{V^2}  \label{eq:106}
\end{equation}

\begin{equation}
\overline{\,\left( \delta S\right) ^2\,}\;\overline{\left( \delta B\right) ^2%
}>k^2T^2H^2\left( \frac{\partial \mu }{\partial T}\right) _\rho ^2
\label{eq:107}
\end{equation}

For a sphere placed in an environment (the corresponding permeabilities
being $\mu _1$ and $\mu _2$), considering also $N$ and $V$ as constants, one
obtains the relations:

\begin{equation}
\overline{\left( \delta T\right) ^2}\;\overline{\left( \delta B_1\right) ^2}%
>0  \label{eq:108}
\end{equation}

\begin{equation}
\overline{\left( \delta T\right) ^2}\;\overline{\,\left( \delta S_1\right) ^2%
}>k^2T^2  \label{eq:109}
\end{equation}

\begin{equation}
\,\overline{\left( \delta B_1\right) ^2}\;\overline{\left( \delta H_1\right)
^2}>\frac{k^2T^2}{V_1^2}\frac{\mu _1^2}{\mu _s^2}  \label{eq:110}
\end{equation}

\begin{equation}
\overline{\,\left( \delta S_1\right) ^2\,}\;\overline{\left( \delta
B_1\right) ^2}>\frac 1{81}k^2T^2B_1^2\mu _s^{\prime \,2}\left( \alpha \frac{%
\partial \mu _2}{\partial T}-\beta \frac{\partial \mu _1}{\partial T}\right)
^2  \label{eq:111}
\end{equation}

\section{SUMMARY, CONCLUSIONS AND CONNECTED REMARKS}

The body of the present paper can be summarized and added with remarks as
follows:

\begin{enumerate}
\item  In the first section it was presented a phenomenological-theoretical
approach of the fluctuations for the macroscopic parameters regarding the
thermodynamic systems taken in the presence of the fields. We started with
the expression of the differential of generalized internal energy. The
respective start was complemented with the consideration that through the
fluctuations the system passes in states which are in the neighbor of a
thermodynamic equilibrium.

\item  In the second section we particularized our approach to the case when
the fields are of electromagnetic nature. We find that the estimators of
fluctuations (i.e. dispersions and correlations) depend on the different
field constraints.

\item  The discrete systems are characterized by the fact that the
fluctuations estimators are functions of both intrinsic quantities of the
system and of the variables regarding the environment.

\item  In the third section we presented a lot of thermodynamic inequalities
which result from the fact that the correlations of the fluctuations
constitute the elements of a non-negatively defined matrix. In their
two-variable versions the respective inequalities are nothing but classical
(non-quantum) analogues of the well known Heisenberg's (''uncertainty'')
relations.

\item  The last specifications from the subsection III.A and B reveal an
interesting feature of the Boltzmann's constant $k$ in the following sense:

\begin{description}
\item  a) The quantities $\overline{\left( \delta A\right) ^2}$ and $%
\overline{\delta A\,\delta B}$ as fluctuation parameters are the estimators
of the level of the stochasticity.

\item  b) The formulas from the mentioned subsections shows the fact that
the respective quantities appears as products of the Boltzmann's constant $k$
with non-stochastic expressions.

\item  c) Then it directly result the idea that $k$ can be regarded as a
generic indicator of thermodynamic stochasticity.
\end{description}

\item  At this point we wish to add the following connected remarks. The
mentioned idea regarding $k$ was firstly promoted in the work \cite{10} of
one of us. In the same work was revealed the similarity of the Boltzmann's
constant $k$ with the Planck's constant $\hbar $. The last one has the
signification of generic indicator of quantum stochasticity. In the cases of
classical (non-quantum) thermodynamical systems respectively of the quantum
microparticles $k$ and $\hbar $ appear independently and singly. Then the
respective systems can be regarded as endowed with an onefold stochasticity.
In the case of the quantum statistical systems $k$ and $\hbar $ appear
together in the expressions of the fluctuation parameters. This means that
such systems are endowed with twofold stochasticity (for more details see 
\cite{10}).
\end{enumerate}

\acknowledgments

\begin{itemize}
\item  We wish to express our deep gratitude to Prof. Y. Zimmels for putting
at our disposition copies of his papers. The respective papers stimulated
our work.

\item  In the end we mention that the researches reported in the present
text take advantage of some facilities of a grant from the Romanian Ministry
of National Education.
\end{itemize}

\begin{table}[tbp] \centering%
\begin{tabular}{|p{1.7in}|p{4.6in}|}
\hline
{\bf Independent variables} & {\bf Inequalities} \\ \hline
$
\begin{array}{l}
S,V,N,B
\end{array}
$ & $\frac{\partial \left( T,-\widehat{p},\widehat{\zeta },VH\right) }{%
\partial \left( S,V,N,B\right) }>0$ \\ \hline
$
\begin{array}{l}
T,V,N,B
\end{array}
$ & $\frac{\partial \left( S,-\widehat{p},\widehat{\zeta },VH\right) }{%
\partial \left( T,V,N,B\right) }>0$ \\ \hline
$
\begin{array}{l}
T,V,N,H \\ 
\left( X_1,X_2,X_3,X_4\right)
\end{array}
$ & $\det \left| \frac{\partial S}{\partial X_b}\delta _{1a}-\frac{\partial 
\widehat{p}}{\partial X_a}\delta _{2b}+\frac{\partial \widehat{\zeta }}{%
\partial X_a}\delta _{3b}+\frac{\partial \left( VH\right) }{\partial X_a}%
\frac{\partial B}{\partial X_b}\right| >0$ \\ \hline
$
\begin{array}{l}
U,V,N,B \\ 
\left( X_1,X_2,X_3,X_4\right)
\end{array}
$ & $\det \left| \frac{\partial T}{\partial X_a}\frac{\partial S}{\partial
X_b}-\frac{\partial \widehat{p}}{\partial X_a}\delta _{2b}+\frac{\partial 
\widehat{\zeta }}{\partial X_a}\delta _{3b}+\frac{\partial \left( VH\right) 
}{\partial X_a}\delta _{4b}\right| >0$ \\ \hline
$
\begin{array}{l}
\frac 1T,V,N,B \\ 
\left( X_1,X_2,X_3,X_4\right)
\end{array}
$ & $\det \left| -T^2\frac{\partial S}{\partial X_b}\delta _{1a}-\frac{%
\partial \widehat{p}}{\partial X_a}\delta _{2b}+\frac{\partial \widehat{%
\zeta }}{\partial X_a}\delta _{3b}+\frac{\partial \left( VH\right) }{%
\partial X_a}\delta _{4b}\right| >0$ \\ \hline
\end{tabular}
\caption{New Thermodynamical Inequalities\label{table1}}%
\end{table}%

\end{document}